\documentclass[aps,twocolumn,pra,showpacs,floatfix]{revtex4}
\usepackage{epsfig}
\usepackage{dcolumn}

\begin{document}

\title{\bf Hyperfine induced electric dipole contributions to the electric octupole and magnetic quadrupole atomic clock transitions}

\author{V. A. Dzuba and V. V. Flambaum}
\affiliation{School of Physics, University of New South Wales, 
Sydney 2052, Australia}

\date{\today}

\begin{abstract}
Hyperfine-induced electric dipole contributions may significantly increase  probabilities of otherwise very weak  electric octupole 
and magnetic quadrupole
 atomic  clock transitions (e.g. transitions between $s$ and $f$ electron orbitals). These transitions can be used for exceptionally 
 accurate atomic clocks, quantum information processing and search for dark matter.
They are very sensitive to new physics beyond the Standard Model, such as  temporal variation of the fine structure constant,
the Lorentz invariance and Einstein equivalence principle violation. 
We formulate conditions under which the hyperfine-induced electric dipole contribution dominates.
Due to the hyperfine quenching the electric octupole clock 
transition in $^{173}$Yb$^+$ is two orders of magnitude stronger  than that in currently used $^{171}$Yb$^+$.
Some enhancement is found in $^{143}$Nd$^{13+}$, $^{149}$Pm$^{14+}$, $^{147}$Sm$^{14+}$, and
$^{147}$Sm$^{15+}$ ions. 

\end{abstract} 
\pacs{06.30.Ft, 06.20.Jr, 31.15.A, 32.30.Jc }
\maketitle

Electric octupole (E3)  and magnetic quadrupole (M2) atomic optical transitions, which correspond to  transitions between 
$s$ and $f$ electron orbitals, 
can be used as optical clocks of exceptionally high 
accuracy~\cite{roberts2000observation,BDF-HCI10,BDFO-hole11,DDF12,HCI-PRL14}.
They also present unique opportunities for fundamental research by being sensitive to new physics beyond
the Standard Model. The transitions are very sensitive to the temporal variation of the fine structure constant $\alpha$ 
($\alpha = e^2/\hbar c$)~\cite{DzuFlaWeb99,DzuFlaWeb99a,
DzuFlaMar03,DzuFla08a,BDF-HCI10,BDFO-hole11,DDF12,HCI-PRL14,Ho14+15}, 
to the local Lorentz invariance (LLI) violation~\cite{dzuba2016strongly}, the effect of dark 
matter~\cite{graham2013new,budker2014proposal,van2015search,
stadnik2014axion,stadnik2014searching,
stadnik2015searching,
stadnik2015can}, etc.
For example, the $4f^{14}6s \ ^2$S$_{1/2} - 4f^{13}6s^2 \ ^2$F$^o_{7/2}$ transition in Yb$^+$ offers
opportunities for frequency measurements with fractional accuracy 
$\sim 10^{-18}$~\cite{roberts2000observation}.
The work is in progress in many laboratories~\cite{tamm2009stray,king2012absolute,rastogi2015design}.
Measuring the ratio of frequency of this transition  to the frequency of the 
$4f^{14}6s \ ^2$S$_{1/2} - 4f^{14}5d \ ^2$D$_{3/2}$ transition in the same ion put the strongest limit on the temporal 
variation of the fine structure constant and (by including the Cs hyperfine transition)  on the proton-to-electron mass ratio~\cite{DzuFlaMar03,DzuFla08a,Tedesco,godun2014frequency,Peik-Yb+14}.
The use of the electric octupole transition in Yb$^+$ for the search of LLI violation may lead to five orders
of magnitude improvement over current best bounds on the LLI violation in the electron-photon 
sector~\cite{dzuba2016strongly}.

Many similar opportunities come with the use of optical transitions in highly-charged ions 
(HCI)~\cite{BDF-HCI10,BDFO-hole11,DDF12,HCI-PRL14,Ho14+15}. For example, the spectrum of the Ir$^{17+}$
ion has been recently measured ~\cite{IrPRL15} with the prospect of using the 
$4f^{13}5s \ ^3$F$^o_{4} - 4f^{12}5s^2 \ ^3$H$_{6}$ transition for the time keeping and fundamental research.

Electrical octupole and magnetic quadrupole transitions are very weak, typical linewidth can be as small as few nHz.
This may lead to certain difficulties in the measurements. In this paper we demonstrate that the electric dipole
transition (E1) induced by hyperfine interaction can be significantly larger than the electric octupole or magnetic 
quadrupole transitions. Therefore, choosing right isotope might be important for the measurements.
 
At least one or more of the following conditions is needed for the domination of the hyperfine-induced 
E1 transitions.
\begin{itemize}
\item The hyperfine mixing is enhanced by a small energy interval. This is may be a fine structure interval
for the ground or clock state. The clock transition in $^{165}$Ho$^{14+}$ ion~\cite{Ho14+15} is an example.
\item 
If  $\Delta J<3$, the magnetic dipole hyperfine interaction (HFI) can contribute to the transition, where  $J$ is the total electron angular momentum. 
Otherwise, an electric quadrupole HFI is needed, which is usually significantly smaller.
For single-valence-electron atoms or ions this would be a $4f_{5/2}$ - $5s$ (or $6s$) transition,
e.g., the $4f_{5/2}$ - $5s$ transition in $^{143}$Nd$^{13+}$. 
For two and more valence electrons, appropriated values of the total electron angular momentum are formed via contribution from all valence electrons.
\item An isotope with a deformed nucleus  has a large electric quadrupole moment. 
This makes the electric quadrupole HFI almost as large as the magnetic dipole HFI. The $4f^{14}6s, \ ^2$S$_{1/2}$
- $4f^{13}6s^2, \ ^2$F$^o_{7/2}$ transition in $^{173}$Yb is an example.
\item  For smaller frequencies all transition rates are smaller. However, the electric octupole transition 
rate depends much stronger on $\omega$ ($\sim \omega^7$) than the electric dipole transition rate
($\sim \omega^3$). Therefore, for a sufficiently  small $\omega$ the electric dipole transition rate always
dominates. The $4f_{5/2}$ - $5s$ transition in $^{149}$Pm$^{14+}$ is an example.
\end{itemize}

The amplitude of the HFI-induced electric dipole transition is given by 
\begin{eqnarray}
&&A_{\rm hfs-E1}(b \rightarrow a) = \label{A-E1-hfs} \\
&&\sum_n \left[ \frac{\langle a |\hat {\rm H}_{\rm hfs}|n\rangle \langle n |\hat {\rm D}|b\rangle} {E_a-E_n}+
\frac{\langle b |\hat {\rm H}_{\rm hfs}|n\rangle \langle n |\hat {\rm D}|a\rangle} {E_b-E_n} \right].
\nonumber
\end{eqnarray}
Here $\hat {\rm H}_{\rm hfs}$ is the Hamiltonian of the magnetic dipole or electric quadrupole HFI, $\hat D$ 
is the electric dipole operator. The detailed expressions with the angular reduction can be found in 
Ref.~\cite{johnson2011hyperfine,porsev2004hyperfine} and in the Appendix.
The corresponding rate for the E1 transition (we use atomic units) is:
\begin{equation}
R_{\rm hfs-E1} = \frac{4}{3}(\omega\alpha)^3\frac{A_{\rm
    hfs-E1}^2}{2F_c+1},
\label{T-E1-hfs}
\end{equation}
where $\omega $  is the frequency of the clock transition, 
$A_{\rm hfs-E1}$ is the amplitude of the transition (\ref{A-E1-hfs}) (the reduced matrix element),
and $F_c$ is the total angular momentum of the clock state including the nuclear spin $I$
($\mathbf{F} = \mathbf{I} + \mathbf{J}$). 
The rate for a E3 transition is  
\begin{equation}
R_{\rm E3} = 0.00169(\omega\alpha)^7\frac{A_{\rm  E3}^2}{2J_c+1},
\label{T-E3}
\end{equation}
where $J_c$ is the total electron angular momentum of the clock state.
The rate for a M2 transition is  
\begin{equation}
R_{\rm M2} = \frac{1}{15}(\omega\alpha)^5\frac{A_{\rm  M2}^2}{2J_c+1},
\label{T-M2}
\end{equation}
where the amplitude of the magnetic quadrupole transition $A_{\rm  M2}$ includes the electron magnetic 
moment $\mu_0$ which is equal in the Gaussian atomic units to $\alpha/2$.

We use the random phase approximation (RPA) to calculate the transition amplitudes and HFI mixing. The RPA
equations for the core states $c$ are
\begin{equation}
	(\hat H^{\rm HF} - \epsilon_c)\delta \psi_c = -(\hat F +\delta V_F)\psi_c,
\label{e:RPA}
\end{equation}
where $\hat H^{\rm HF}$ is the relativistic Hartree-Fock (HF) Hamiltonian, $\psi_c$ and $\epsilon_c$
are single-electron HF state in the core  and its energy, ($\hat H^{\rm HF} -\epsilon_c)\psi_c = 0$, 
$\delta \psi_c$ is the correction to the core state $c$ induced by an external field, $\hat F$ is the 
operator of the external filed, and $\delta V_F$ is the correction to the self-consistent HF potential
induced by the external field via corrections to all core states. Equations (\ref{e:RPA}) are solved self-consistently
for all states in the core. Amplitude of the transition between valence states $v$ and $w$ 
(or mixing of these states) in the RPA approximation is given by the matrix element
\begin{equation}
A_{wv} = \langle w|\hat F + \delta V| v \rangle.
\label{e:me}
\end{equation}
In present work we consider five different external field operators. These include the electric dipole and octupole
operators, the  magnetic quadrupole operator, the magnetic dipole and electric quadrupole HFI operators. 
Expressions for the single-electron matrix elements for each of these operators are presented in the Appendix.

In case of single-valence-electron atom or ion the amplitude (\ref{A-E1-hfs}) can be reduced to
\begin{eqnarray}
&&A_{\rm hfs-E1}(b \rightarrow a) = \label{e1-hfs} \\
&&\langle \delta \psi_a|\hat d + \delta V_d| \psi_b \rangle +
      \langle \psi_a|\hat d + \delta V_d| \delta \psi_b \rangle, \nonumber
\end{eqnarray}
where $\delta \psi_a$ and $\delta \psi_b$ are corrections to valence states $a$ and $b$ induced by HFI,
$\hat d = -e{\mathbf r}$ is the single-electron electric dipole operator, $\delta V_d$ is the correction to
the HF potential of the core induced by the electric field of the photon.

In the case of several valence electrons we use the configuration interaction (CI) 
technique~\cite{DzuFla08,DzuFla08a}
to construct many-electron states $a$, $b$ and $n$ in (\ref{A-E1-hfs}). Then we perform direct summation 
over excited states $n$. The summation is truncated at sufficiently high states so that the tail contribution is 
reasonably small. 

\begin{table*}
\caption{Hyperfine structure of the ground and clock states of the ions considered in this work.
$I$ is the nuclear spin, $\mu$ is the nuclear magnetic dipole moment in nuclear magnetons, $Q$ is the 
nuclear electric quadrupole moment in barns ($10^{-28}$m$^2$), $A$ is the magnetic dipole hfs constant
(in MHz), $B$ is the electric quadrupole hfs constant (in MHz). A comparison with an experiment is given for Yb$^+$.}
\label{t:hfs}
\begin{ruledtabular}
\begin{tabular}{lld c ll cc ll cc}
\multicolumn{1}{c}{Ion} & 
\multicolumn{1}{c}{$I$} & 
\multicolumn{1}{c}{$\mu$} & 
\multicolumn{1}{c}{$Q$} & 
\multicolumn{2}{c}{Ground} & 
\multicolumn{1}{c}{$A$} & 
\multicolumn{1}{c}{$B$} & 
\multicolumn{2}{c}{Clock} & 
\multicolumn{1}{c}{$A$} & 
\multicolumn{1}{c}{$B$} \\ 
\multicolumn{1}{c}{Isotope} & &
\multicolumn{1}{c}{$\mu_N$} & 
\multicolumn{1}{c}{b} & 
\multicolumn{2}{c}{State} & 
\multicolumn{1}{c}{MHz} & 
\multicolumn{1}{c}{MHz} & 
\multicolumn{2}{c}{State} & 
\multicolumn{1}{c}{MHz} & 
\multicolumn{1}{c}{MHz} \\ 
\hline
$^{143}$Nd$^{13+}$  & 7/2 & -1.08 & -0.630(60) & $5s$ & $^2$S$_{1/2}$ & -38200 &  0 & $4f$ & $^2$F$^o_{5/2}$  & -333 & -833 \\

$^{149}$Pm$^{14+}$ & 7/2 &   \pm 3.3  &  no data    & $5s$ & $^2$S$_{1/2}$ & $\pm$130500 &  0 & $4f$ & $^2$F$^o_{5/2}$ & $\pm$1162 & 1530\footnotemark[1]\\

$^{147}$Sm$^{15+}$ &  7/2 & -0.813 & -0.259(26) & $4f$ & $^2$F$^o_{5/2}$   & -318 &-443& $5s$ & $^2$S$_{1/2}$ & -34800 & 0 \\

$^{147}$Sm$^{14+}$ &  7/2 & -0.813 & -0.259(26) & $4f^2$ & $^3$H$_4$   &-320 & 3728 & $5s4f$ & $^3$F$^o_2$ & 5222 & 2833  \\

$^{193}$Ir$^{17+}$ &  3/2 & 0.1591 & 0.751(9) & $4f^{13}5s $ & $ ^3$F$^o_4$ & 5180 &-3218 & $4f^{12}5s^2 $ & $ ^3$H$_6$ & 176 & -3332  \\

$^{171}$Yb$^{+}$ &  1/2 & 0.4919 & 0 & $4f^{14}6s $ & $^2$S$_{1/2}$   &11600& 0 & $4f^{13}6s^2 $ & $ ^3$F$^o_{7/2}$ & 871 & 0  \\
$^{171}$Yb$^{+}$ &  \multicolumn{2}{c}{Experiment\footnotemark[2]} &&&   & 12645 &  &   &   &  905.0(5) &   \\

$^{173}$Yb$^{+}$ &  5/2 & -0.6776 & 2.800(4) &  $4f^{14}6s $ & $ ^2$S$_{1/2}$   &-3200& 0 & $4f^{13}6s^2 $ & $ ^3$F$^o_{7/2}$ & -240 &  -4762 \\
$^{173}$Yb$^{+}$ &  \multicolumn{2}{c}{Experiment\footnotemark[3]} &&&   &-3497.5(6) &  &   &   &  &   \\
\end{tabular}
\end{ruledtabular}
\footnotetext[1]{Assuming $Q=1$b.}
\footnotetext[2]{Ref.~\cite{taylor1999measurement}.}
\footnotetext[3]{Ref.~\cite{maartensson1994isotope}.}
\end{table*}

\begin{table}
\caption{Experimental or theoretical frequencies of the clock transitions ($\omega$) and theoretical 
rates of the spontaneous decay of the clock state due to the electric octupole (E3), 
or magnetic quadrupole (M2) transitions to the ground state for the ions of Table~\ref{t:hfs}.
The experimental rate for Yb$^+$ is $0.59(+1.21/-0.38)\times 10^{-8} \rm{s}^{-1}$~\cite{roberts2000observation}.} 
\label{t:E3M2}
\begin{ruledtabular}
\begin{tabular}{l l l l l rrr}
\multicolumn{1}{c}{Ion} & 
\multicolumn{2}{c}{Ground} & 
\multicolumn{2}{c}{Clock} & 
\multicolumn{1}{c}{$\omega$} & 
\multicolumn{1}{c}{$R_{\rm E3}$} & 
\multicolumn{1}{c}{$R_{\rm M2}$} \\
&\multicolumn{2}{c}{State} & 
\multicolumn{2}{c}{State} & 
\multicolumn{1}{c}{cm$^{-1}$} & 
\multicolumn{1}{c}{s$^{-1}$} & 
\multicolumn{1}{c}{s$^{-1}$} \\ 
\hline
Nd$^{13+}$ & $5s$ &  $^2$S$_{1/2}$  & $4f$ & $^2$F$^o_{5/2}$ &55870\footnotemark[1] &  8.0[-7] & 4.6[-11]  \\

Pm$^{14+}$ & $5s$ & $^2$S$_{1/2}$ & $4f$ &$^2$F$^o_{5/2}$ & 3228\footnotemark[2] & 1.2[-14] & 2.5[-17]  \\

Sm$^{15+}$ &  $4f$  & $^2$F$^o_{5/2}$ & $5s$ & $^2$S$_{1/2}$ & 60384\footnotemark[1] & 2.2[-6] & 1.1[-8]  \\

Sm$^{14+}$ &  $4f^2$ & $^3$H$_4$   & $5s4f$ & $^3$F$^o_2$ & 2172\footnotemark[3] & 1.5[-16] & 1.5[-14]  \\
 Ir$^{17+}$   & $4f^{13}5s $ & $ ^3$F$^o_4$ &  $4f^{12}5s^2 $ & $ ^3$H$_6$ & 37423\footnotemark[4] & 7.9[-9] & 3.0[-9]  \\

Yb$^{+}$ &  $4f^{14}6s $ & $^2$S$_{1/2}$  & $4f^{13}6s^2 $ & $ ^3$F$^o_{7/2}$ & 21419\footnotemark[5] & 1.0[-8]& 0 \\
\end{tabular}
\end{ruledtabular}
\footnotetext[1]{Expt.,~\cite{SugarNd13Sm15}.}
\footnotetext[2]{Theory, this work.}
\footnotetext[3]{Theory,~\cite{Cd-likeHCI14}.}
\footnotetext[4]{Expt.~\cite{IrPRL15}.}
\footnotetext[5]{Expt.~\cite{NIST}.}
\end{table}

\begin{table}
\caption{The rate of the HFI-induced electric dipole transitions between hyperfine components of the ground and
clock states of the  ions for Table~\ref{t:hfs}.}
\label{t:R}
\begin{ruledtabular}
\begin{tabular}{lc ddr}
\multicolumn{1}{c}{Ion} & 
\multicolumn{1}{c}{Clock} & &&
\multicolumn{1}{c}{$R_{\rm E1-HFI}$} \\ 
\multicolumn{1}{c}{Isotope} & 
\multicolumn{1}{c}{Transition} & 
\multicolumn{1}{c}{$F_g$} & 
\multicolumn{1}{c}{$F_c$} & 
\multicolumn{1}{c}{s$^{-1}$} \\ 
\hline
$^{143}$Nd$^{13+}$ & $^2$S$_{1/2} - ^2$F$^o_{5/2}$ & 3.0 & 2.0 & 7.2[-8] \\
                                  && 3.0 & 3.0 & 1.1[-7] \\
                                   && 3.0 & 4.0 & 8.2[-8] \\
                                  && 4.0 & 3.0 & 3.7[-8] \\
                                   && 4.0 & 4.0 & 1.1[-7] \\
                                   && 4.0 & 5.0 & 1.7[-7] \\

$^{149}$Pm$^{14+}$ & $^2$S$_{1/2} - ^2$F$^o_{5/2}$& 3.0 & 2.0 & 4.8[-11] \\
                                   && 3.0 & 3.0 & 7.4[-11] \\
                                   && 3.0 & 4.0 & 5.5[-11] \\
                                   && 4.0 & 3.0 & 2.5[-11] \\
                                   && 4.0 & 4.0 & 7.6[-11] \\
                                   && 4.0 & 5.0 & 1.1[-10] \\

$^{147}$Sm$^{15+}$ & $^2$F$^o_{5/2} - ^2$S$_{1/2}$ & 2.0 & 3.0 & 5.0[-9] \\
                                   && 3.0 & 3.0 & 1.1[-8] \\
                                   && 3.0 & 4.0 & 2.9[-9] \\
                                   && 4.0 & 3.0 & 1.0[-8] \\
                                   && 4.0 & 4.0 & 1.1[-8] \\
                                   && 5.0 & 4.0 & 2.0[-8] \\

$^{147}$Sm$^{14+}$ & $^3$H$_{4} - ^3$F$^o_{2}$&  0.5 &  1.5 & 6.1[-10]  \\
& &  1.5 &  1.5 & 2.8[-9] \\
& &  1.5 &  2.5 & 1.2[-9] \\
& &  2.5 &  1.5 & 2.9[-9] \\
& &  2.5 &  2.5 & 5.9[-9] \\
 &&  2.5 &  3.5 & 1.4[-9] \\
 &&  3.5 &  2.5 & 7.8[-9] \\
 &&  3.5 &  3.5 & 7.9[-9] \\
 &&  3.5 &  4.5 & 1.0[-9] \\
 &&  4.5 &  3.5 & 1.4[-8] \\
& &  4.5 &  4.5 & 7.5[-9] \\
 &&  4.5 &  5.5 & 4.1[-10] \\
& &  5.5 &  4.5 & 1.9[-8] \\
&&  5.5 &  5.5 & 4.3[-9] \\
 &&  6.5 &  5.5 & 1.7[-8] \\

$^{193}$Ir$^{17+}$ & $^3$F$^o_{4} - ^3$H$_{6}$ &  3.5 & 4.5 & 5.0[-9] \\
&&  4.5 & 4.5 & 4.7[-10] \\
&&  5.5 & 4.5 & 6.7[-12] \\
&&  4.5 & 5.5 & 3.8[-10] \\
&&  5.5 & 5.5 & 2.2[-11] \\
&&  5.5 & 6.5 & 4.0[-9] \\

$^{173}$Yb$^{+}$ & $^2$S$_{1/2} - ^3$F$^o_{7/2}$ &  2.0 & 1.0 &  3.3[-7] \\
&&  2.0 & 2.0 &  1.8[-6] \\
&&  3.0 & 2.0 &  5.1[-7] \\
&&  2.0 & 3.0 &  2.5[-6] \\
&&  3.0 & 3.0 &  3.2[-6] \\
&&  3.0 & 4.0 &  6.5[-6] \\
\end{tabular}
\end{ruledtabular}
\end{table}

A byproduct of this work is the calculated hyperfine structure of the ground and clock states 
of specific isotopes of several ions considered for the search of 
$\alpha$-variation~\cite{DDF12,HCI-PRL14,Ag-likeHCI14,IrPRL15} and the Lorentz invariance 
violation~\cite{dzuba2016strongly}.
This includes, e.g. the Ir$^{17+}$ ion for which first measurements of the spectra were 
recently reported~\cite{IrPRL15} confirming the predicted $5s$-$4f$ level crossing. 
The $^{171}$Yb$^+$ ions are used by several 
experimental groups as prospective optical clocks of an exceptional high 
accuracy~\cite{king2012absolute,Peik-Yb+14,godun2014frequency,rastogi2015design,leute2015frequency}. 
Current best limits on the
temporal variation of the fine structure constant come from the comparison of the frequencies of the
$4f^{14}6s \ ^2$S$_{1/2}$ - $4f^{13}6s^2 \ ^3$F$_{2/2}$ E3 and $4f^{12}6s \ ^2$S$_{1/2}$ -
$4f^{14}5d \ ^2$D$_{3/2}$ E2 transitions~\cite{DzuFlaMar03,DzuFla08a,godun2014frequency,Peik-Yb+14}.  
The results for the hyperfine structure are presented in Table \ref{t:hfs}. Apart from getting new 
data, the comparison with the experiment for $^{171}$Yb$^+$ and $^{173}$Yb$^+$ tests the accuracy of
the calculations. It indicates that the theoretical uncertainty is smaller than 10\%.

Table~\ref{t:E3M2} presents calculated E3 and M2 transition rates for the clock states of the ions.
They are to be compared with the HFI-induced rates in Table~\ref{t:R}. The calculated E3 transition rate for Yb$^+$
is within error bars of the experimental value~\cite{roberts2000observation}. 
The E3 transition rate is larger than the M2 rate in all
ions except for the Sm$^{14+}$ ion, where it is small due to the small frequency.

The calculated HFI-induced E1 transition rates are presented in Table~\ref{t:R}.
They are larger than the  E3 or M2 rates for the $^{143}$Nd$^{13+}$, $^{149}$Pm$^{14+}$,
$^{147}$Sm$^{14+}$, and $^{173}$Yb$^{+}$ ions (see Table~\ref{t:E3M2}).
The most interesting case is probably the $^{173}$Yb$^{+}$ ion. This is due
to its importance as one of the most accurate optical clocks, and study of the temporal
variation of fine structure constant and the Lorentz invariance violation as discussed above.
Currently used $^{171}$Yb$^{+}$ ion has a very weak clock transition rate, about two
orders of magnitude weaker than the rate  in the $^{173}$Yb$^{+}$ ion calculated in the present work. This is due to the HFI-induced
E1 transition in $^{173}$Yb$^{+}$ which is relatively large due to the large contribution
from the electric quadrupole hyperfine mixing. The $^{171}$Yb$^{13+}$ isotope
has small nuclear spin ($I=1/2$) and no nuclear electric quadrupole moment. 

The main reason for using odd isotope for the E3 clock transition in Yb$^+$ is the
possibility to eliminate linear Zeeman shift by using states with $M=0$~\cite{roberts2000observation} 
($M$ is the projection of the total angular momentum of the atom, including nuclear spin).
This works for both $^{171}$Yb$^{+}$ and $^{173}$Yb$^{+}$. Note that the electric quadrupole
shift due to gradients of electric field can also be eliminated in both isotopes
by considering states with $F=3$, $M=\pm 2$ ($\mathbf{F}=\mathbf{J}+\mathbf{I}$, $M$ is 
projection of $\mathbf{F}$, the energy shift $\Delta E \sim 3M^2 -F(F+1)$ vanishes for $F=3$, $M=\pm 2$). In this case the linear Zeeman
shift can be eliminated by averaging the measurements involving states with $M=+2$ and 
$M=-2$.

\acknowledgments

The work was supported in part by the Australian Research Council.

\appendix
\section{Matrix elements}

The matrix element of the electromagnetic transition amplitude between many-electron states with definite values
of the electron total angular momentum $J$ and the total atomic angular momentum $F$ ($\mathbf{F}=\mathbf(J)+\mathbf{I}$, 
where $I$ is the nuclear spin) is given by
\begin{eqnarray}
&&\langle J_a,F_a|| A_k||J_b,F_b \rangle = (-1)^{I+F_b+J_a+k}\times \label{e:Ak} \\
&& \sqrt{(2F_a+1)(2F_b+1)} \left\{ \begin{array}{ccc} J_b & J_a & k \\ F_a & F_b & I \end{array} \right\} \langle J_a|| A_k || J_b \rangle,
\nonumber
\end{eqnarray}
where $k$ is the operator rank ($k=1$ for E1 and M1, $k=2$ for E2 and M2, etc.).

The matrix element of the magnetic dipole HFI is
\begin{eqnarray}
&&\langle J_a,F|| \hat H_A||J_b,F \rangle = (-1)^{I+F+J_b}\times \label{e:HA} \\
&& \mu \sqrt{\frac{(2I+1)(I+1)}{I}}\left\{ \begin{array}{ccc} I & J_b & F \\ J_a & I & 1 \end{array} \right\} 
\langle J_a|| \hat H_A || J_b \rangle.
\nonumber
\end{eqnarray}
Here $\mu$ is the nuclear magnetic dipole moment in nuclear magnetons.

The matrix element of the electric quadrupole HFI is
\begin{eqnarray}
&&\langle J_a,F|| \hat H_B||J_b,F \rangle = \frac{Q}{2}  (-1)^{I+F+J_b}\times \label{e:HB} \\
&&  \sqrt{\frac{(I+1)(2I+1)(2I+3)}{I(2I-1)}} \left\{ \begin{array}{ccc} I & J_b & F \\ J_a & I & 2 \end{array} \right\} 
\langle J_a|| \hat H_B || J_b \rangle.
\nonumber
\end{eqnarray}
Here $Q$ is the nuclear electric quadrupole moment in atomic units.

In expressions (\ref{e:Ak}), (\ref{e:HA}) and (\ref{e:HB}) the matrix element  $\langle J_a|| \hat  H || J_b \rangle$
is the matrix element of a one-body operator between many-electron CI wave functions of valence electrons.
The standard CI technique is used to reduce theses matrix elements to the single-electron matrix elements.

We use single-electron wave functions in a form
\begin{equation}
    \psi(r)_{njlm}=\frac{1}{r}\left(\begin {array}{c}
    f_{v}(r)\Omega(\mathbf{n})_{\mathit{jlm}}  \\[0.2ex]
    i\alpha g_{v}(r)  \widetilde{ \Omega}(\mathbf{n})_{\mathit{jlm}}
    \end{array} \right),
\label{psi}
\end{equation}
where $n$ is the principal quantum number and the index $v$
replaces the three-number set $n,j,l$; $\alpha$ is the fine structure
constant. 

The matrix elements of the electric multipole transitions are ($\alpha\omega r \ll 1$)
\begin{eqnarray}
&&\langle a || E_k|| b \rangle = \langle \kappa_a||C_k||\kappa_b \rangle  \times \nonumber \\
&&\int \left(f_a(r)f_b(r)+\alpha^2g_a(r)g_b(r)\right) r^kdr
\label{e:Ek}
\end{eqnarray}

The matrix elements of the magnetic quadrupole transitions are 
\begin{eqnarray}
&&\langle a || M_2|| b \rangle = \langle -\kappa_a||C_2||\kappa_b \rangle  (\kappa_a+\kappa_b) \times \nonumber \\
&&\frac{2\alpha}{3} \int \left(f_a(r)g_b(r)+g_a(r)f_b(r)\right) r^2dr
\label{e:Mk}
\end{eqnarray}

The matrix elements of the magnetic dipole HFI  are 
\begin{eqnarray}
&&\langle a || \hat H_{A}|| b \rangle = -\langle -\kappa_a||C_1||\kappa_b \rangle  (\kappa_a+\kappa_b) \times \nonumber \\
&& \frac{\alpha^2}{2} \frac{m_e}{m_p} \int \left(f_a(r)g_b(r)+g_a(r)f_b(r)\right) \frac{r}{r_>^3}dr
\label{e:Ahfs}
\end{eqnarray}

The matrix elements of the electric quadrupole HFI are 
\begin{eqnarray}
&&\langle a || \hat H_B|| b \rangle = \langle \kappa_a||C_2||\kappa_b \rangle  \times \nonumber \\
&&\int \left(f_a(r)f_b(r)+\alpha^2g_a(r)g_b(r)\right) \frac{r}{r_>^4}dr
\label{e:Bhfs}
\end{eqnarray}
In (\ref{e:Ahfs}) and (\ref{e:Bhfs}) $r_>=\max(r,r_N)$, where $r_N$ is the nuclear radius.

The reduced matrix element of the spherical harmonic $C_k$ is
\begin{eqnarray}
&\langle \kappa_a||C_k||\kappa_b \rangle = 
(-1)^{j_b+1/2}\sqrt{(2j_a+1)(2j_b+1)} \nonumber \\
& \times \xi(l_a+l_b+k)\left( \begin{array}{ccc} j_b & j_a & k \\
-1/2 & 1/2 & 0 \end{array} \right) . \label{eq:ck} \\
& \xi(x)= \left\{ \begin{array}{ccccc} 1, & {\rm if} &x& {\rm is}&{\rm even} \\
0, & {\rm if} &x& {\rm is}&{\rm odd} \end{array} \right. \nonumber
\end{eqnarray}

\bibliographystyle{apsrev}

\begin{thebibliography}{35}
\expandafter\ifx\csname natexlab\endcsname\relax\def\natexlab#1{#1}\fi
\expandafter\ifx\csname bibnamefont\endcsname\relax
  \def\bibnamefont#1{#1}\fi
\expandafter\ifx\csname bibfnamefont\endcsname\relax
  \def\bibfnamefont#1{#1}\fi
\expandafter\ifx\csname citenamefont\endcsname\relax
  \def\citenamefont#1{#1}\fi
\expandafter\ifx\csname url\endcsname\relax
  \def\url#1{\texttt{#1}}\fi
\expandafter\ifx\csname urlprefix\endcsname\relax\def\urlprefix{URL }\fi
\providecommand{\bibinfo}[2]{#2}
\providecommand{\eprint}[2][]{\url{#2}}

\bibitem[{\citenamefont{Roberts et~al.}(2000)\citenamefont{Roberts, Taylor,
  Barwood, Rowley, and Gill}}]{roberts2000observation}
\bibinfo{author}{\bibfnamefont{M.}~\bibnamefont{Roberts}},
  \bibinfo{author}{\bibfnamefont{P.}~\bibnamefont{Taylor}},
  \bibinfo{author}{\bibfnamefont{G.~P.} \bibnamefont{Barwood}},
  \bibinfo{author}{\bibfnamefont{W.~R.~C.} \bibnamefont{Rowley}},
  \bibnamefont{and} \bibinfo{author}{\bibfnamefont{P.}~\bibnamefont{Gill}},
  \bibinfo{journal}{Phys. Rev. A} \textbf{\bibinfo{volume}{62}},
  \bibinfo{pages}{020501} (\bibinfo{year}{2000}).

\bibitem[{\citenamefont{Berengut et~al.}(2010)\citenamefont{Berengut, Dzuba,
  and Flambaum}}]{BDF-HCI10}
\bibinfo{author}{\bibfnamefont{J.~C.} \bibnamefont{Berengut}},
  \bibinfo{author}{\bibfnamefont{V.~A.} \bibnamefont{Dzuba}}, \bibnamefont{and}
  \bibinfo{author}{\bibfnamefont{V.~V.} \bibnamefont{Flambaum}},
  \bibinfo{journal}{Phys. Rev. Lett.} \textbf{\bibinfo{volume}{105}},
  \bibinfo{pages}{120801} (\bibinfo{year}{2010}).

\bibitem[{\citenamefont{Berengut et~al.}(2011)\citenamefont{Berengut, Dzuba,
  Flambaum, and Ong}}]{BDFO-hole11}
\bibinfo{author}{\bibfnamefont{J.~C.} \bibnamefont{Berengut}},
  \bibinfo{author}{\bibfnamefont{V.~A.} \bibnamefont{Dzuba}},
  \bibinfo{author}{\bibfnamefont{V.~V.} \bibnamefont{Flambaum}},
  \bibnamefont{and} \bibinfo{author}{\bibfnamefont{A.}~\bibnamefont{Ong}},
  \bibinfo{journal}{Phys. Rev. Lett.} \textbf{\bibinfo{volume}{106}},
  \bibinfo{pages}{210802} (\bibinfo{year}{2011}).

\bibitem[{\citenamefont{Dzuba et~al.}(2012)\citenamefont{Dzuba, Derevianko, and
  Flambaum}}]{DDF12}
\bibinfo{author}{\bibfnamefont{V.~A.} \bibnamefont{Dzuba}},
  \bibinfo{author}{\bibfnamefont{A.}~\bibnamefont{Derevianko}},
  \bibnamefont{and} \bibinfo{author}{\bibfnamefont{V.~V.}
  \bibnamefont{Flambaum}}, \bibinfo{journal}{Phys. Rev. A}
  \textbf{\bibinfo{volume}{86}}, \bibinfo{pages}{054502}
  (\bibinfo{year}{2012}).

\bibitem[{\citenamefont{Safronova
  et~al.}(2014{\natexlab{a}})\citenamefont{Safronova, Dzuba, Flambaum,
  Safronova, Porsev, and Kozlov}}]{HCI-PRL14}
\bibinfo{author}{\bibfnamefont{M.~S.} \bibnamefont{Safronova}},
  \bibinfo{author}{\bibfnamefont{V.~A.} \bibnamefont{Dzuba}},
  \bibinfo{author}{\bibfnamefont{V.~V.} \bibnamefont{Flambaum}},
  \bibinfo{author}{\bibfnamefont{U.~I.} \bibnamefont{Safronova}},
  \bibinfo{author}{\bibfnamefont{S.~G.} \bibnamefont{Porsev}},
  \bibnamefont{and} \bibinfo{author}{\bibfnamefont{M.~G.}
  \bibnamefont{Kozlov}}, \bibinfo{journal}{Phys. Rev. Lett.}
  \textbf{\bibinfo{volume}{113}}, \bibinfo{pages}{030801}
  (\bibinfo{year}{2014}{\natexlab{a}}).

\bibitem[{\citenamefont{Dzuba et~al.}(1999{\natexlab{a}})\citenamefont{Dzuba,
  Flambaum, and Webb}}]{DzuFlaWeb99}
\bibinfo{author}{\bibfnamefont{V.~A.} \bibnamefont{Dzuba}},
  \bibinfo{author}{\bibfnamefont{V.~V.} \bibnamefont{Flambaum}},
  \bibnamefont{and} \bibinfo{author}{\bibfnamefont{J.~K.} \bibnamefont{Webb}},
  \bibinfo{journal}{Phys. Rev. Lett.} \textbf{\bibinfo{volume}{82}},
  \bibinfo{pages}{888} (\bibinfo{year}{1999}{\natexlab{a}}).

\bibitem[{\citenamefont{Dzuba et~al.}(1999{\natexlab{b}})\citenamefont{Dzuba,
  Flambaum, and Webb}}]{DzuFlaWeb99a}
\bibinfo{author}{\bibfnamefont{V.~A.} \bibnamefont{Dzuba}},
  \bibinfo{author}{\bibfnamefont{V.~V.} \bibnamefont{Flambaum}},
  \bibnamefont{and} \bibinfo{author}{\bibfnamefont{J.~K.} \bibnamefont{Webb}},
  \bibinfo{journal}{Phys. Rev. A} \textbf{\bibinfo{volume}{59}},
  \bibinfo{pages}{230} (\bibinfo{year}{1999}{\natexlab{b}}).

\bibitem[{\citenamefont{Dzuba et~al.}(2003)\citenamefont{Dzuba, Flambaum, and
  Marchenko}}]{DzuFlaMar03}
\bibinfo{author}{\bibfnamefont{V.~A.} \bibnamefont{Dzuba}},
  \bibinfo{author}{\bibfnamefont{V.~V.} \bibnamefont{Flambaum}},
  \bibnamefont{and} \bibinfo{author}{\bibfnamefont{M.~V.}
  \bibnamefont{Marchenko}}, \bibinfo{journal}{Phys. Rev. A}
  \textbf{\bibinfo{volume}{68}}, \bibinfo{pages}{022506}
  (\bibinfo{year}{2003}).

\bibitem[{\citenamefont{Dzuba and Flambaum}(2008{\natexlab{a}})}]{DzuFla08a}
\bibinfo{author}{\bibfnamefont{V.~A.} \bibnamefont{Dzuba}} \bibnamefont{and}
  \bibinfo{author}{\bibfnamefont{V.~V.} \bibnamefont{Flambaum}},
  \bibinfo{journal}{Phys. Rev. A} \textbf{\bibinfo{volume}{77}},
  \bibinfo{pages}{012515} (\bibinfo{year}{2008}{\natexlab{a}}).

\bibitem[{\citenamefont{Dzuba et~al.}(2015)\citenamefont{Dzuba, Flambaum, and
  Katori}}]{Ho14+15}
\bibinfo{author}{\bibfnamefont{V.~A.} \bibnamefont{Dzuba}},
  \bibinfo{author}{\bibfnamefont{V.~V.} \bibnamefont{Flambaum}},
  \bibnamefont{and} \bibinfo{author}{\bibfnamefont{H.}~\bibnamefont{Katori}},
  \bibinfo{journal}{Phys. Rev. A} \textbf{\bibinfo{volume}{91}},
  \bibinfo{pages}{022119} (\bibinfo{year}{2015}).

\bibitem[{\citenamefont{Dzuba et~al.}(2016)\citenamefont{Dzuba, Flambaum,
  Safronova, Porsev, Pruttivarasin, Hohensee, and
  H{\"a}ffner}}]{dzuba2016strongly}
\bibinfo{author}{\bibfnamefont{V.~A.} \bibnamefont{Dzuba}},
  \bibinfo{author}{\bibfnamefont{V.~V.} \bibnamefont{Flambaum}},
  \bibinfo{author}{\bibfnamefont{M.~S.} \bibnamefont{Safronova}},
  \bibinfo{author}{\bibfnamefont{S.~G.} \bibnamefont{Porsev}},
  \bibinfo{author}{\bibfnamefont{T.}~\bibnamefont{Pruttivarasin}},
  \bibinfo{author}{\bibfnamefont{M.~A.} \bibnamefont{Hohensee}},
  \bibnamefont{and}
  \bibinfo{author}{\bibfnamefont{H.}~\bibnamefont{H{\"a}ffner}},
  \bibinfo{journal}{Nature Physics}  (\bibinfo{year}{2016}).

\bibitem[{\citenamefont{Graham and Rajendran}(2013)}]{graham2013new}
\bibinfo{author}{\bibfnamefont{P.~W.} \bibnamefont{Graham}} \bibnamefont{and}
  \bibinfo{author}{\bibfnamefont{S.}~\bibnamefont{Rajendran}},
  \bibinfo{journal}{Phys. Rev. D} \textbf{\bibinfo{volume}{88}},
  \bibinfo{pages}{035023} (\bibinfo{year}{2013}).

\bibitem[{\citenamefont{Budker et~al.}(2014)\citenamefont{Budker, Graham,
  Ledbetter, Rajendran, and Sushkov}}]{budker2014proposal}
\bibinfo{author}{\bibfnamefont{D.}~\bibnamefont{Budker}},
  \bibinfo{author}{\bibfnamefont{P.~W.} \bibnamefont{Graham}},
  \bibinfo{author}{\bibfnamefont{M.}~\bibnamefont{Ledbetter}},
  \bibinfo{author}{\bibfnamefont{S.}~\bibnamefont{Rajendran}},
  \bibnamefont{and} \bibinfo{author}{\bibfnamefont{A.~O.}
  \bibnamefont{Sushkov}}, \bibinfo{journal}{Phys. Rev. X}
  \textbf{\bibinfo{volume}{4}}, \bibinfo{pages}{021030} (\bibinfo{year}{2014}).

\bibitem[{\citenamefont{Van~Tilburg et~al.}(2015)\citenamefont{Van~Tilburg,
  Leefer, Bougas, and Budker}}]{van2015search}
\bibinfo{author}{\bibfnamefont{K.}~\bibnamefont{Van~Tilburg}},
  \bibinfo{author}{\bibfnamefont{N.}~\bibnamefont{Leefer}},
  \bibinfo{author}{\bibfnamefont{L.}~\bibnamefont{Bougas}}, \bibnamefont{and}
  \bibinfo{author}{\bibfnamefont{D.}~\bibnamefont{Budker}},
  \bibinfo{journal}{Phys. Rev. Lett.} \textbf{\bibinfo{volume}{115}},
  \bibinfo{pages}{011802} (\bibinfo{year}{2015}).

\bibitem[{\citenamefont{Stadnik and
  Flambaum}(2014{\natexlab{a}})}]{stadnik2014axion}
\bibinfo{author}{\bibfnamefont{Y.~V.} \bibnamefont{Stadnik}} \bibnamefont{and}
  \bibinfo{author}{\bibfnamefont{V.~V.} \bibnamefont{Flambaum}},
  \bibinfo{journal}{Phys. Rev. D} \textbf{\bibinfo{volume}{89}},
  \bibinfo{pages}{043522} (\bibinfo{year}{2014}{\natexlab{a}}).

\bibitem[{\citenamefont{Stadnik and
  Flambaum}(2014{\natexlab{b}})}]{stadnik2014searching}
\bibinfo{author}{\bibfnamefont{Y.~V.} \bibnamefont{Stadnik}} \bibnamefont{and}
  \bibinfo{author}{\bibfnamefont{V.~V.} \bibnamefont{Flambaum}},
  \bibinfo{journal}{Phys. Rev. Lett.} \textbf{\bibinfo{volume}{113}},
  \bibinfo{pages}{151301} (\bibinfo{year}{2014}{\natexlab{b}}).

\bibitem[{\citenamefont{Stadnik and
  Flambaum}(2015{\natexlab{a}})}]{stadnik2015searching}
\bibinfo{author}{\bibfnamefont{Y.~V.} \bibnamefont{Stadnik}} \bibnamefont{and}
  \bibinfo{author}{\bibfnamefont{V.~V.} \bibnamefont{Flambaum}},
  \bibinfo{journal}{Phys. Rev. Lett.} \textbf{\bibinfo{volume}{114}},
  \bibinfo{pages}{161301} (\bibinfo{year}{2015}{\natexlab{a}}).

\bibitem[{\citenamefont{Stadnik and
  Flambaum}(2015{\natexlab{b}})}]{stadnik2015can}
\bibinfo{author}{\bibfnamefont{Y.~V.} \bibnamefont{Stadnik}} \bibnamefont{and}
  \bibinfo{author}{\bibfnamefont{V.~V.} \bibnamefont{Flambaum}},
  \bibinfo{journal}{Phys. Rev. Lett.} \textbf{\bibinfo{volume}{115}},
  \bibinfo{pages}{201301} (\bibinfo{year}{2015}{\natexlab{b}}).

\bibitem[{\citenamefont{Tamm et~al.}(2009)\citenamefont{Tamm, Weyers,
  Lipphardt, and Peik}}]{tamm2009stray}
\bibinfo{author}{\bibfnamefont{C.}~\bibnamefont{Tamm}},
  \bibinfo{author}{\bibfnamefont{S.}~\bibnamefont{Weyers}},
  \bibinfo{author}{\bibfnamefont{B.}~\bibnamefont{Lipphardt}},
  \bibnamefont{and} \bibinfo{author}{\bibfnamefont{E.}~\bibnamefont{Peik}},
  \bibinfo{journal}{Phys. Rev. A} \textbf{\bibinfo{volume}{80}},
  \bibinfo{pages}{043403} (\bibinfo{year}{2009}).

\bibitem[{\citenamefont{King et~al.}(2012)\citenamefont{King, Godun, Webster,
  Margolis, Johnson, Szymaniec, Baird, and Gill}}]{king2012absolute}
\bibinfo{author}{\bibfnamefont{S.~A.} \bibnamefont{King}},
  \bibinfo{author}{\bibfnamefont{R.~M.} \bibnamefont{Godun}},
  \bibinfo{author}{\bibfnamefont{S.~A.} \bibnamefont{Webster}},
  \bibinfo{author}{\bibfnamefont{H.~S.} \bibnamefont{Margolis}},
  \bibinfo{author}{\bibfnamefont{L.~A.~M.} \bibnamefont{Johnson}},
  \bibinfo{author}{\bibfnamefont{K.}~\bibnamefont{Szymaniec}},
  \bibinfo{author}{\bibfnamefont{P.~E.~G.} \bibnamefont{Baird}},
  \bibnamefont{and} \bibinfo{author}{\bibfnamefont{P.}~\bibnamefont{Gill}},
  \bibinfo{journal}{New J. Phys.} \textbf{\bibinfo{volume}{14}},
  \bibinfo{pages}{013045} (\bibinfo{year}{2012}).

\bibitem[{\citenamefont{Rastogi et~al.}(2015)\citenamefont{Rastogi, Batra, Roy,
  Thangjam, Kalsi, Panja, and De}}]{rastogi2015design}
\bibinfo{author}{\bibfnamefont{A.}~\bibnamefont{Rastogi}},
  \bibinfo{author}{\bibfnamefont{N.}~\bibnamefont{Batra}},
  \bibinfo{author}{\bibfnamefont{A.}~\bibnamefont{Roy}},
  \bibinfo{author}{\bibfnamefont{J.}~\bibnamefont{Thangjam}},
  \bibinfo{author}{\bibfnamefont{V.~P.~S.} \bibnamefont{Kalsi}},
  \bibinfo{author}{\bibfnamefont{S.}~\bibnamefont{Panja}}, \bibnamefont{and}
  \bibinfo{author}{\bibfnamefont{S.}~\bibnamefont{De}},
  \bibinfo{journal}{MAPAN} \textbf{\bibinfo{volume}{30}}, \bibinfo{pages}{169}
  (\bibinfo{year}{2015}).

\bibitem[{\citenamefont{Flambaum and Tedesco}(2006)}]{Tedesco}
\bibinfo{author}{\bibfnamefont{V.~V.} \bibnamefont{Flambaum}} \bibnamefont{and}
  \bibinfo{author}{\bibfnamefont{A.~F.} \bibnamefont{Tedesco}},
  \bibinfo{journal}{Phys. Rev. C} \textbf{\bibinfo{volume}{73}},
  \bibinfo{pages}{055501} (\bibinfo{year}{2006}).

\bibitem[{\citenamefont{Godun et~al.}(2014)\citenamefont{Godun, Nisbet-Jones,
  Jones, King, Johnson, Margolis, Szymaniec, Lea, Bongs, and
  Gill}}]{godun2014frequency}
\bibinfo{author}{\bibfnamefont{R.~M.} \bibnamefont{Godun}},
  \bibinfo{author}{\bibfnamefont{P.~B.~R.} \bibnamefont{Nisbet-Jones}},
  \bibinfo{author}{\bibfnamefont{J.~M.} \bibnamefont{Jones}},
  \bibinfo{author}{\bibfnamefont{S.~A.} \bibnamefont{King}},
  \bibinfo{author}{\bibfnamefont{L.~A.~M.} \bibnamefont{Johnson}},
  \bibinfo{author}{\bibfnamefont{H.~S.} \bibnamefont{Margolis}},
  \bibinfo{author}{\bibfnamefont{K.}~\bibnamefont{Szymaniec}},
  \bibinfo{author}{\bibfnamefont{S.~N.} \bibnamefont{Lea}},
  \bibinfo{author}{\bibfnamefont{K.}~\bibnamefont{Bongs}}, \bibnamefont{and}
  \bibinfo{author}{\bibfnamefont{P.}~\bibnamefont{Gill}},
  \bibinfo{journal}{Phys. Rev. Lett.} \textbf{\bibinfo{volume}{113}},
  \bibinfo{pages}{210801} (\bibinfo{year}{2014}).

\bibitem[{\citenamefont{Huntemann et~al.}(2014)\citenamefont{Huntemann,
  Lipphardt, Tamm, Gerginov, Weyers, and Peik}}]{Peik-Yb+14}
\bibinfo{author}{\bibfnamefont{N.}~\bibnamefont{Huntemann}},
  \bibinfo{author}{\bibfnamefont{B.}~\bibnamefont{Lipphardt}},
  \bibinfo{author}{\bibfnamefont{C.}~\bibnamefont{Tamm}},
  \bibinfo{author}{\bibfnamefont{V.}~\bibnamefont{Gerginov}},
  \bibinfo{author}{\bibfnamefont{S.}~\bibnamefont{Weyers}}, \bibnamefont{and}
  \bibinfo{author}{\bibfnamefont{E.}~\bibnamefont{Peik}},
  \bibinfo{journal}{Phys. Rev. Lett.} \textbf{\bibinfo{volume}{113}},
  \bibinfo{pages}{210802} (\bibinfo{year}{2014}).

\bibitem[{\citenamefont{Windberger et~al.}(2015)\citenamefont{Windberger,
  López-Urrutia, Bekker, Oreshkinan, Berengut, Bock, Borschevsky, Dzuba, Eliav,
  Harman et~al.}}]{IrPRL15}
\bibinfo{author}{\bibfnamefont{A.}~\bibnamefont{Windberger}},
  \bibinfo{author}{\bibfnamefont{J.~R.~C.} \bibnamefont{López-Urrutia}},
  \bibinfo{author}{\bibfnamefont{H.}~\bibnamefont{Bekker}},
  \bibinfo{author}{\bibfnamefont{N.}~\bibnamefont{Oreshkinan}},
  \bibinfo{author}{\bibfnamefont{J.}~\bibnamefont{Berengut}},
  \bibinfo{author}{\bibfnamefont{V.}~\bibnamefont{Bock}},
  \bibinfo{author}{\bibfnamefont{A.}~\bibnamefont{Borschevsky}},
  \bibinfo{author}{\bibfnamefont{V.~A.} \bibnamefont{Dzuba}},
  \bibinfo{author}{\bibfnamefont{E.}~\bibnamefont{Eliav}},
  \bibinfo{author}{\bibfnamefont{Z.}~\bibnamefont{Harman}},
  \bibnamefont{et~al.}, \bibinfo{journal}{Phys. Rev. Lett.}
  \textbf{\bibinfo{volume}{114}}, \bibinfo{pages}{150801}
  (\bibinfo{year}{2015}).

\bibitem[{\citenamefont{Johnson}(2011)}]{johnson2011hyperfine}
\bibinfo{author}{\bibfnamefont{W.~R.} \bibnamefont{Johnson}},
  \bibinfo{journal}{Canadian J. Phys.} \textbf{\bibinfo{volume}{89}},
  \bibinfo{pages}{429} (\bibinfo{year}{2011}).

\bibitem[{\citenamefont{Porsev and Derevianko}(2004)}]{porsev2004hyperfine}
\bibinfo{author}{\bibfnamefont{S.~G.} \bibnamefont{Porsev}} \bibnamefont{and}
  \bibinfo{author}{\bibfnamefont{A.}~\bibnamefont{Derevianko}},
  \bibinfo{journal}{Phys. Rev. A} \textbf{\bibinfo{volume}{69}},
  \bibinfo{pages}{042506} (\bibinfo{year}{2004}).

\bibitem[{\citenamefont{Dzuba and Flambaum}(2008{\natexlab{b}})}]{DzuFla08}
\bibinfo{author}{\bibfnamefont{V.~A.} \bibnamefont{Dzuba}} \bibnamefont{and}
  \bibinfo{author}{\bibfnamefont{V.~V.} \bibnamefont{Flambaum}},
  \bibinfo{journal}{Phys. Rev. A} \textbf{\bibinfo{volume}{77}},
  \bibinfo{pages}{012514} (\bibinfo{year}{2008}{\natexlab{b}}).

\bibitem[{\citenamefont{Taylor et~al.}(1999)\citenamefont{Taylor, Roberts,
  Macfarlane, Barwood, Rowley, and Gill}}]{taylor1999measurement}
\bibinfo{author}{\bibfnamefont{P.}~\bibnamefont{Taylor}},
  \bibinfo{author}{\bibfnamefont{M.}~\bibnamefont{Roberts}},
  \bibinfo{author}{\bibfnamefont{G.~M.} \bibnamefont{Macfarlane}},
  \bibinfo{author}{\bibfnamefont{G.~P.} \bibnamefont{Barwood}},
  \bibinfo{author}{\bibfnamefont{W.~R.~C.} \bibnamefont{Rowley}},
  \bibnamefont{and} \bibinfo{author}{\bibfnamefont{P.}~\bibnamefont{Gill}},
  \bibinfo{journal}{Phys. Rev. A} \textbf{\bibinfo{volume}{60}},
  \bibinfo{pages}{2829} (\bibinfo{year}{1999}).

\bibitem[{\citenamefont{M{\aa}rtensson-Pendrill
  et~al.}(1994)\citenamefont{M{\aa}rtensson-Pendrill, Gough, and
  Hannaford}}]{maartensson1994isotope}
\bibinfo{author}{\bibfnamefont{A.-M.} \bibnamefont{M{\aa}rtensson-Pendrill}},
  \bibinfo{author}{\bibfnamefont{D.~S.} \bibnamefont{Gough}}, \bibnamefont{and}
  \bibinfo{author}{\bibfnamefont{P.}~\bibnamefont{Hannaford}},
  \bibinfo{journal}{Phys. Rev. A} \textbf{\bibinfo{volume}{49}},
  \bibinfo{pages}{3351} (\bibinfo{year}{1994}).

\bibitem[{\citenamefont{Sugar and Kaufman}(1981)}]{SugarNd13Sm15}
\bibinfo{author}{\bibfnamefont{J.}~\bibnamefont{Sugar}} \bibnamefont{and}
  \bibinfo{author}{\bibfnamefont{V.}~\bibnamefont{Kaufman}},
  \bibinfo{journal}{Phys. Scr.} \textbf{\bibinfo{volume}{24}},
  \bibinfo{pages}{742} (\bibinfo{year}{1981}).

\bibitem[{\citenamefont{Safronova
  et~al.}(2014{\natexlab{b}})\citenamefont{Safronova, Dzuba, Flambaum,
  Safronova, Porsev, and Kozlov}}]{Cd-likeHCI14}
\bibinfo{author}{\bibfnamefont{M.~S.} \bibnamefont{Safronova}},
  \bibinfo{author}{\bibfnamefont{V.~A.} \bibnamefont{Dzuba}},
  \bibinfo{author}{\bibfnamefont{V.~V.} \bibnamefont{Flambaum}},
  \bibinfo{author}{\bibfnamefont{U.~I.} \bibnamefont{Safronova}},
  \bibinfo{author}{\bibfnamefont{S.~G.} \bibnamefont{Porsev}},
  \bibnamefont{and} \bibinfo{author}{\bibfnamefont{M.~G.}
  \bibnamefont{Kozlov}}, \bibinfo{journal}{Phys. Rev. A}
  \textbf{\bibinfo{volume}{90}}, \bibinfo{pages}{052509}
  (\bibinfo{year}{2014}{\natexlab{b}}).

\bibitem[{\citenamefont{Kramida et~al.}(2015)\citenamefont{Kramida,
  {Yu.~Ralchenko}, Reader, and {and NIST ASD Team}}}]{NIST}
\bibinfo{author}{\bibfnamefont{A.}~\bibnamefont{Kramida}},
  \bibinfo{author}{\bibnamefont{{Yu.~Ralchenko}}},
  \bibinfo{author}{\bibfnamefont{J.}~\bibnamefont{Reader}}, \bibnamefont{and}
  \bibinfo{author}{\bibnamefont{{and NIST ASD Team}}},
  \bibinfo{howpublished}{{NIST Atomic Spectra Database (ver. 5.3), [Online].
  Available: {\tt{http://physics.nist.gov/asd}} [2016, January 11]. National
  Institute of Standards and Technology, Gaithersburg, MD.}}
  (\bibinfo{year}{2015}).

\bibitem[{\citenamefont{Safronova
  et~al.}(2014{\natexlab{c}})\citenamefont{Safronova, Dzuba, Flambaum,
  Safronova, Porsev, and Kozlov}}]{Ag-likeHCI14}
\bibinfo{author}{\bibfnamefont{M.~S.} \bibnamefont{Safronova}},
  \bibinfo{author}{\bibfnamefont{V.~A.} \bibnamefont{Dzuba}},
  \bibinfo{author}{\bibfnamefont{V.~V.} \bibnamefont{Flambaum}},
  \bibinfo{author}{\bibfnamefont{U.~I.} \bibnamefont{Safronova}},
  \bibinfo{author}{\bibfnamefont{S.~G.} \bibnamefont{Porsev}},
  \bibnamefont{and} \bibinfo{author}{\bibfnamefont{M.~G.}
  \bibnamefont{Kozlov}}, \bibinfo{journal}{Phys. Rev. A}
  \textbf{\bibinfo{volume}{90}}, \bibinfo{pages}{042513}
  (\bibinfo{year}{2014}{\natexlab{c}}).

\bibitem[{\citenamefont{Leute et~al.}(2015)\citenamefont{Leute, Huntemann,
  Lipphardt, Tamm, Nisbet-Jones, King, Godun, Jones, Margolis, Whibberley
  et~al.}}]{leute2015frequency}
\bibinfo{author}{\bibfnamefont{J.}~\bibnamefont{Leute}},
  \bibinfo{author}{\bibfnamefont{N.}~\bibnamefont{Huntemann}},
  \bibinfo{author}{\bibfnamefont{B.}~\bibnamefont{Lipphardt}},
  \bibinfo{author}{\bibfnamefont{C.}~\bibnamefont{Tamm}},
  \bibinfo{author}{\bibfnamefont{P.~B.~R.} \bibnamefont{Nisbet-Jones}},
  \bibinfo{author}{\bibfnamefont{S.~A.} \bibnamefont{King}},
  \bibinfo{author}{\bibfnamefont{R.~M.} \bibnamefont{Godun}},
  \bibinfo{author}{\bibfnamefont{J.~M.} \bibnamefont{Jones}},
  \bibinfo{author}{\bibfnamefont{H.~S.} \bibnamefont{Margolis}},
  \bibinfo{author}{\bibfnamefont{P.~B.} \bibnamefont{Whibberley}},
  \bibnamefont{et~al.}, \bibinfo{journal}{arXiv preprint arXiv:1507.04754}
  (\bibinfo{year}{2015}).

\end{thebibliography}

\end{document}